\documentclass[aps,twocolumn,nobibnotes,pra]{revtex4}
\usepackage{amsmath,epsfig,graphicx,amssymb}
\usepackage{bm}%
\expandafter\ifx\csname package@font\endcsname\relax\else
 \expandafter\expandafter
 \expandafter\usepackage
 \expandafter\expandafter
 \expandafter{\csname package@font\endcsname}%
\fi

\begin{document}
\title{Spin-flip M1 giant resonance as a challenge for Skyrme forces}
\author{V.O. Nesterenko$^1$, J. Kvasil$^2$,  P. Vesely$^{2,3}$, W. Kleinig$^{1,4}$,
P.-G. Reinhard$^5$, and V.Yu. Ponomarev $^6$}
\affiliation{$^{1}$ BLTP, Joint Institute for Nuclear Research, 141980, Dubna,
Moscow region, Russia; nester@theor.jinr.ru}
\affiliation{$^{2}$ Institute of Particle and Nuclear
Physics, Charles University,
CZ-18000, Praha, Czech Republic;\\
kvasil@ipnp.troja.mff.cuni.cz; vesely@ipnp.troja.mff.cuni.cz}
\affiliation
{$^{3}$ Department of Physics, P.O. Box 35 (YFL), University of
Jyvaskyla, 40014, Jyvaskyla, Finland}
\affiliation{$^{4}$
Technische Universit\"at Dresden, Inst. f\"ur Analysis, D-01062, Dresden,
Germany}
\affiliation{$^{5}$
Institut f\"ur Theoretische Physik II, Universit\"at Erlangen, D-91058,
Erlangen, Germany}
\affiliation{$^6$ Institut f\"ur
Kernphysik, Technische Universit\"at Darmstadt, D-64289, Darmstadt, Germany}
\date{\today}

\begin{abstract}
Despite a great success of the Skyrme mean-field approach in exploration of
nuclear dynamics, it  seems to fail in description of the spin-flip M1 giant
resonance. The results for different Skyrme parameterizations are contradictory
and poorly agree with experiment. In particular, there is no parameterization
which simultaneously describes the one-peak gross structure of M1 strength in
doubly magic nuclei and two-peak structure in heavy deformed nuclei. The reason
of this mismatch could lie in an unsatisfactory treatment of spin correlations
and spin-orbit interaction. We discuss the present status of the problem and
possible ways of its solution. In particular, we inspect i) the interplay of
the collective shift and spin-orbit splitting, ii) the isovector M1 response
versus isospin-mixed responses, and iii) the role of tensor and isovector
spin-orbit interaction.
\end{abstract}

\maketitle

\section{Introduction}

The spin-flip M1 giant resonance (M1GR) was a subject of intensive theoretical
and experimental studies during the past decades
\cite{Harakeh_book_01,Speth_91,Osterfeld_92}. The resonance is known to be a
major source of knowledge on spin correlations. Besides, it strongly depends on
the spin-orbit splitting and so can serve as a robust test of the spin-orbit
interaction. The M1GR was widely explored within various empirical microscopic
models, see e.g. \cite{Kam_83,Pon_87,Zaw_90,Coster_90}, which allowed to
clarify its main features. Meanwhile, the nuclear density functional theory
(DFT) has been developed. It provided elaborate self-consistent methods
(Skyrme, Gogny, relativistic) with high descriptive power
\cite{Ben,Vre05aR,Sto07aR}. Hence, it is now desirable to study the M1GR in
this context. Until recently, most of the DFT applications to nuclear dynamics
were concentrated on electric modes and Gamow-Teller (GT) resonance
\cite{Ben,Vre05aR,Sto07aR}, while much less work was done for magnetic
excitations. At the same time, the exploration of magnetic modes in general and
spin-flip M1GR in particular could be extremely useful to clarify the spin and
spin-orbit correlations in the nuclear density functionals. This holds
especially for the Skyrme and Gogny functionals where, unlike the relativistic
models, the spin-orbit interaction is an independent part of the modeling.
Further, magnetic modes allow to explore the spin terms in spin-saturated
even-even nuclei, where they cannot be fitted by the ground state properties.
The M1GR can help to clarify still vague role of the tensor forces
\cite{Les_PRC_07_tensor,Colo_Sagawa_08,Marg_Sagawa_JPG_09,Sagawa_lanl_09,vesely_PRC_09}.
And, last but not least, the spin-flip M1GR is a counterpart of the GT
resonance which is of great current interest in connection with astrophysical
problems \cite{Vre05aR,Sto07aR}. So, a satisfactory treatment of M1GR is
relevant for the DFT description of the GT resonance as well.

In this paper, we will concentrate on the exploration of the spin-flip M1GR
within the Skyrme-Hartree-Fock (SHF) approach \cite{Skyrme,Vau,Engel_75}.  To
the best of our knowledge, there are only few early SHF studies of this
resonance \cite{Sarriguren_M1,Hilton_98} but even they are not fully
consistent. Indeed, the study \cite{Sarriguren_M1} exploits a hybrid model with
a partial implementation of SHF in the Landau-Migdal formulation while the work
\cite{Hilton_98} uses early Skyrme forces and, what is crucial, omits the
important spin-density correlations. Only recently the first fully
self-consistent systematic SHF investigation of the spin-flip M1GR was
performed \cite{vesely_PRC_09}.  The calculations were done within the
separable Random-Phase-Approximation (SRPA) model
\cite{nest_PRC_02,nest_PRC_06,nest_PRC_08} extended to magnetic excitations
\cite{vesely_PRC_09,Petr_PhD}. The resonance was explored in spherical nuclei
$^{48}$Ca and $^{208}$Pb and deformed nuclei $^{158}$Gd and $^{238}$U.  Eight
different Skyrme parameterizations were considered and it was shown that none
of them is able to describe simultaneously the one-peak structure of the
resonance in doubly-magic nuclei together with the two-peak structure in
deformed nuclei. The main reason of the failure seems to be in a poor
description of the interplay between the collective shifts (caused by
spin-density correlations) and spin-orbit splitting in the static mean field.
Obviously, this failure of Skyrme forces is also an alarming message
for SHF investigations of the GT resonance.

It is also worth mentioning the very recent SHF study \cite{Sagawa_lanl_09} where,
in accordance with results \cite{vesely_PRC_09}, a considerable influence of tensor
forces on the spin-flip M1GR in spherical nuclei was found. Hence,
the tensor forces become indeed an important, though still not well understood,
factor in the exploration of this resonance.

Altogether, one may state that the M1GR is still a challenge for SHF
and leaves very serious open problems. A discussion of these problems
is the main scope of the present paper. We will also discuss the
possibility to use the M1GR for testing the spin,
spin-orbit and tensor terms in the Skyrme functional. The interplay
of these terms is rather involved and makes the problem indeed
demanding. We
will discuss the present status of the studies, scrutinize some
particular important points (isovector character of M1GR and its
manifestation in experiment, contributions of the tensor and isovector
spin-orbit terms, etc) and sketch the possible
ways of the further progress.

The exploration is performed within the self-consistent separable
Random-Phase-Approximation (SRPA) model
\cite{vesely_PRC_09,nest_IJMPE_09,nest_PRC_02,nest_PRC_06,Petr_PhD}
based on the Skyrme functional \cite{Skyrme,Vau,Engel_75}. The model
was shown to be an effective and accurate tool for systematic study of
multipole electric giant resonances
\cite{nest_PRC_02,nest_PRC_06,nest_PRC_08}. Recently, it was extended
and applied to magnetic excitations
\cite{vesely_PRC_09,nest_IJMPE_09,Petr_PhD}.

The paper is outlined as follows. In Sec. 2, the SRPA model is sketched.  In
Sec. 3, the present status of the SHF description of M1GR and related
difficulties are summarized.  In Sec. 4, the isospin character of the measured
and computed M1GR responses is discussed as a possible source of the
discrepancies. In Sec. 5, the possible important role of the tensor and
isovector spin-orbit terms is considered. In Sec. 6, the conclusions are drawn.

\section{Model and calculation scheme}
\label{sec:model}

SRPA is a fully self-consistent DFT model where both the static
mean field and residual interaction are derived from the Skyrme functional
\cite{Skyrme,Vau,Engel_75}. The SRPA residual interaction includes all
contributions from the Skyrme functional as well as the Coulomb (direct and
exchange) and pairing (at BCS level) terms.  The self-consistent factorization
of the residual interaction in SRPA considerably reduces the computational
expense while maintaining a high accuracy. This makes the model very suitable
for systematic studies. The model was firstly derived and widely used for
electric excitations \cite{nest_PRC_02,nest_PRC_06,nest_PRC_08}. Recently it
was extended to magnetic modes \cite{vesely_PRC_09,nest_IJMPE_09,Petr_PhD}.

Starting point is the Skyrme energy-density functional \cite{Ben,Sto07aR}
\begin{eqnarray}
  \mathcal{H}_\mathrm{Sk}
  &=&
  \frac{b_0}{2} \rho^2- \frac{b'_0}{2} \sum_q\rho_{q}^2
 + b_1 (\rho \tau - \textbf{j}^2)
 - b'_1 \sum_q(\rho_q \tau_q - \textbf{j}^2_q)
\nonumber
\\
 &&
 - \frac{b_2}{2} \rho\Delta \rho
 + \frac{b'_2}{2} \sum_q \rho_q \Delta \rho_q
  + \frac{b_3}{3} \rho^{\alpha+2}
  - \frac{b'_3}{3} \rho^{\alpha} \sum_q \rho^2_q
\nonumber
\\
 &&
 - b_4 (\rho \nabla\textbf{J}\!+
        \!(\nabla\!\times\!\textbf{j})\!\cdot\!\textbf{s})
 - b'_4 \sum_q (\rho_q \nabla\textbf{J}_q\!
 +\!(\nabla\!\times\!\textbf{j}_q)\!\cdot \!\textbf{s}_q)
\nonumber
\\
  &&
  + \frac{\tilde{b}_0}{2} \textbf{s}^2
  - \frac{\tilde{b}'_0}{2} \sum_q \textbf{s}_{q}^2
-\frac{\tilde{b}_2}{2} \textbf{s} \!\cdot\!
  \Delta \textbf{s} + \frac{\tilde{b}'_2}{2}
  \sum_q \textbf{s}_q \!\cdot\!\Delta \textbf{s}_q
\nonumber
\\
  &&
+ \frac{\tilde{b}_3}{3} \rho^{\alpha} \textbf{s}^2
- \frac{\tilde{b}'_3}{3} \rho^{\alpha} \sum_q \textbf{s}^2_q
\nonumber
\\
 &&
  +\gamma_\mathrm{T}(\tilde{b}_1
   (\textbf{s}\!\cdot\!\textbf{T}\!-\!\textbf{J}^2)
  + \tilde{b}'_1
   \sum_q (\textbf{s}_q\!\cdot\!\textbf{T}_q
    \!-\!\textbf{J}_q^2))
\label{eq:skyrme_funct}
\end{eqnarray}
where $b_i$, $b'_i$, $\tilde{b}_i$, $\tilde{b}'_i$ are the force parameters.
The functional involves time-even (nucleon $\rho_q$, kinetic-energy $\tau_q$,
spin-orbit $\textbf{J}_q$) and time-odd (current $\textbf{j}_{ q}$, spin
$\textbf{s}_q$, vector kinetic-energy $\textbf{T}_q$) densities where $q$
denotes protons and neutrons. The total densities, like $\rho = \rho_p +
\rho_n$, are without the index. The contributions with $b_i$ and $b'_i$
(i=0,1,2,3,4) represent the standard terms responsible for the ground state properties
and electric excitations of even-even nuclei \cite{Ben,Sto07aR}. In traditional
SHF functionals, the isovector spin-orbit interaction is linked to the
isoscalar one by $b'_4=b_4$. The tensor spin-orbit terms
$\propto\tilde{b}_1,\tilde{b}'_1$ are often omitted. In Eq.
(\ref{eq:skyrme_funct}) they can be switched by the parameter
$\gamma_\mathrm{T}$.  The spin terms with $\tilde{b}_i, \tilde{b}'_i$ become
relevant only for odd nuclei and magnetic modes in even-even nuclei.  Though
$\tilde{b}_i, \tilde{b}'_i$ may be uniquely determined as functions of
${b}_i,{b}'_i$ \cite{Sto07aR}, their values were not yet well tested by nuclear
data and so are usually considered as free parameters. Just these spin terms
are of paramount importance for the spin-flip M1GR.

SRPA is a {\it fully} self-consistent model as its residual interaction includes all
the terms following from the initial Skyrme functional. For magnetic modes,
these terms are determined through the second functional derivatives
\begin{equation}\label{spin_contr}
\frac{\delta^2 E}{\delta \textbf{j}_{q'} \delta \textbf{s}_{q}} \; ,
\frac{\delta^2 E}{\delta \textbf{s}_{q'} \delta \textbf{s}_{q}} \; ,
\frac{\delta^2 E}{\delta \textbf{J}_{q'} \delta \textbf{J}_{q}} \; ,
\frac{\delta^2 E}{\delta \textbf{T}_{q'} \delta \textbf{s}_{q}} \; .
\end{equation}

The pairing comes through the functional $V_\mathrm{pair}=1/2\sum_q
G_q\chi_q^{\mbox{}}\chi^*_q$ where $\chi_q$ is the pairing density and $G_q$ is
the pairing strength \cite{Sto07aR}. In the present study, pairing is included
at the BCS level through the quasiparticle energies and Bogoliubov's
coefficients.  Unlike the case of the scissors mode,  a possible violation of
the particle number conservation  is not critical for spin-flip M1GR with its
rather high energy.  Anyway, a better pairing description within SRPA is in
progress.

The spectral distribution of the spin-flip M1 mode with $K^{\pi}=1^+$
is given by the strength function
\begin{equation}
  S(M1 ; \omega) = \sum_{\nu \ne 0}
  |\langle\Psi_\nu|\hat{M}|\Psi_0\rangle |^2
  \zeta(\omega - \omega_{\nu})
\label{eq:strength_function}
\end{equation}
where $\Psi_0$ is the ground state, $\nu$ runs over the RPA
$K^{\pi}=1^+$ states with energies $\omega_{\nu}$ and wave functions
$\Psi_\nu$. Further, $ \zeta(\omega - \omega_{\nu}) = \Delta
/[2\pi[(\omega- \omega_{\nu})^2+\frac{\Delta^2}{4}]]$ is a Lorentz
weight with the averaging parameter $\Delta$=1 MeV.  Such averaging
serves to simulate broadening effects beyond SRPA (escape widths,
coupling with complex configurations) and the width $\Delta$ is chosen
to be optimal for the comparison with experiment. The strength function
(\ref{eq:strength_function}) is computed directly, i.e. without
calculation of RPA states $\nu$, which reduces the
computation expense even more.

The operator of spin-flip M1 transition in
(\ref{eq:strength_function}) reads \cite{Osterfeld_92}
\begin{eqnarray}\label{M1_oper}
\hat{M}&=&\mu_B \sqrt{\frac{3}{8\pi}}
[g^{p}_s\sum_{i=1}^Z  {\hat s}_i
 + g^{n}_s\sum_{i=1}^N  {\hat s}_i]
 \\
&=&\mu_B\sqrt{\frac{3}{8\pi}}\sum_{i=1}^A
[\frac{1}{2} g^{0}_s - \tau_3 g^{1}_s] {\hat s}_i
\end{eqnarray}
where ${\hat s}_i$ is the spin operator, $g^{p}_s = 5.58 \varsigma_p$ and
$g^{n}_s = - 3.82 \varsigma_n$ are proton and neutron spin $g$-factors,
$g^{0}_s=g^{p}_s + g^{n}_s=1.35$ and $g^{1}_s=g^{p}_s - g^{n}_s=6.24$ are
isoscalar (T=0) and isovector (T=1) spin $g$-factors, the isospin $\tau_3$ is
-1/2 for protons and 1/2 for neutrons. All the $g$-factors are quenched by
$\varsigma_p$=0.68 and $\varsigma_n$=0.64. Note that $g^{1}_s >> g^{0}_s$ which
shows the predominantly isovector character of the spin-flip M1 resonance. As
we are interested in the spin-flip transitions, the orbital part in
(\ref{M1_oper}) is omitted. Note that in the experimental data
\cite{exp_Gd_U,exp_208Pb} used later for the comparison, the orbital
contribution is strongly suppressed.

SRPA calculations employ a coordinate-space grid with a mesh size
of 0.7 fm.  For deformed nuclei, cylindrical coordinates are used and
the equilibrium quadrupole deformation is found by minimization of the
total energy \cite{nest_PRC_06,nest_PRC_08}.  The single-particle
states are taken into account from the bottom of the potential well up
to +20 MeV. In the heaviest nucleus under consideration, $^{238}U$,
this gives $\sim$17000 two-quasiparticle (2qp) $K^{\pi}=1^+$ pairs
with the excitation energies up to 50-70 MeV.  More details of the
SRPA formalism and calculation scheme can be found in
\cite{vesely_PRC_09,nest_IJMPE_09,Petr_PhD}.

\section{Present status of the problem}

In the present study, the M1 strength is considered in spherical
($^{208}$Pb) and deformed ($^{158}$Gd and $^{238}$U) nuclei.  A representative
set of eight SHF parameterizations is used: SkT6 \cite{skt6}, SkO \cite{sko},
SkO' \cite{sko}, SG2 \cite{sg2}, SkM* \cite{skms}, SLy6 \cite{sly46}, SkI4
\cite{ski3}, and SV-bas \cite{svbas}. They exhibit a variety of effective
masses (from $m^*/m$=1 in SkT6 down to 0.65 in SkI4) and other nuclear matter
characteristics. Some of the forces (SLy6) were found best in the description
of E1(T=1) GR \cite{nest_PRC_06,nest_PRC_08,nest_ijmp}. Others were used in
studies of Gamow-Teller strength (SG2, SkO')
\cite{sg2,Bender_PRC_02_GT,Fracasso_PRC_07_GT,Sarrin_NPA_01_GT} or
peculiarities of spin-orbit splitting (SkI4) \cite{ski3}.  The forces SkT6, SG2
and SkO' involve the tensor spin-orbit term added with (SkO') and
without (SkT6, SG2) refitting the Skyrme parameters. SV-bas is one of the
latest SHF parameterizations \cite{svbas} where the spin-orbit isovector
interaction is varied freely by setting $b'_4 \ne b_4$.
\begin{figure}
\begin{center}
\includegraphics[height=8.5cm,width=8cm]{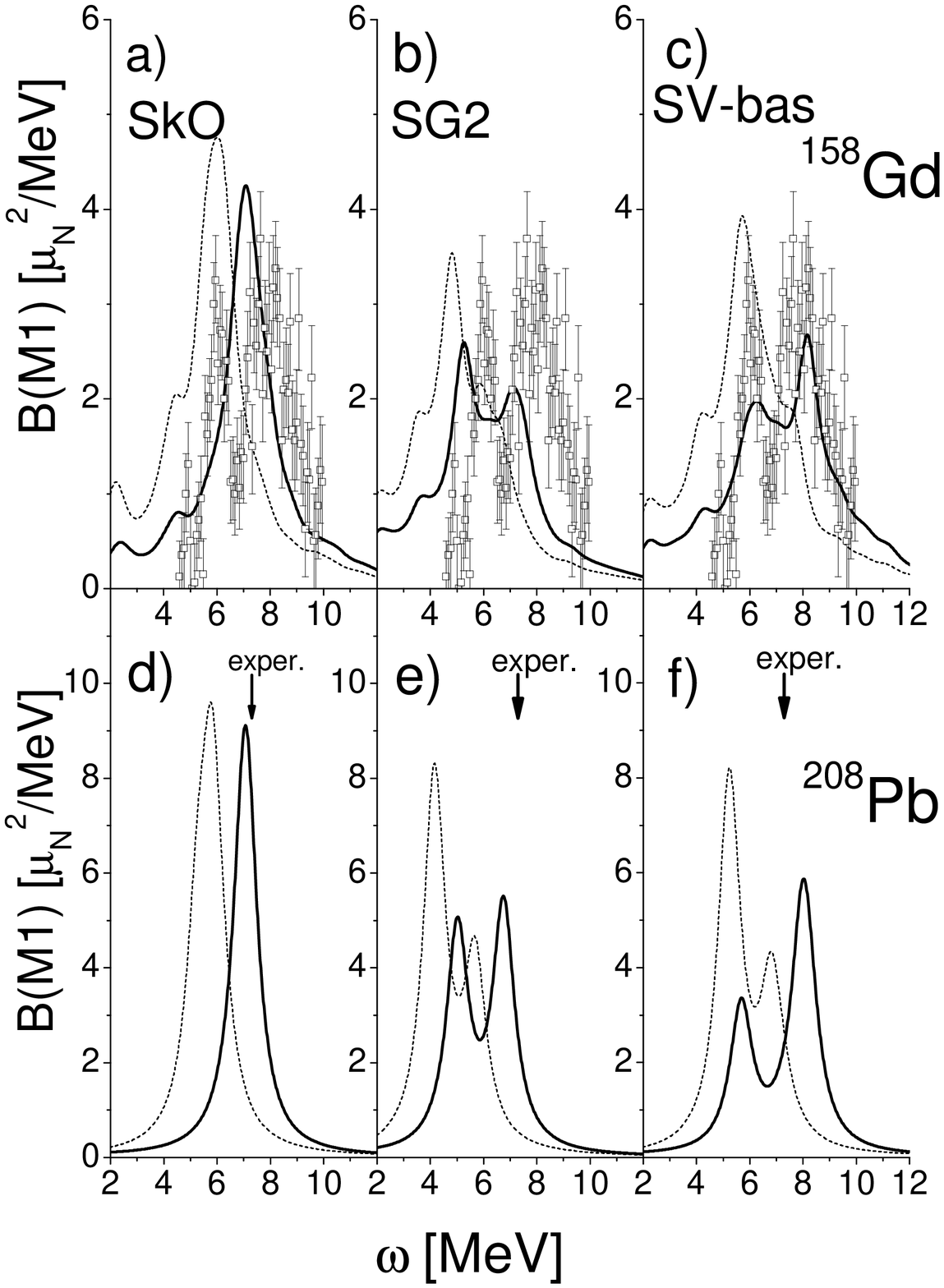}
\caption{ \label{fig:fig1_BM1} The unperturbed (short-dash curve) and SRPA
(solid curve) M1 strength in $^{158}$Gd and $^{208}$Pb for the forces SkO, SG2,
and SV-bas. The experimental data are given by boxes with bars for $^{158}$Gd
\protect\cite{exp_Gd_U} and vertical arrows for $^{208}$Pb
\protect\cite{exp_208Pb}. The strength is smoothed by the Lorentz weight with
$\Delta$=1 MeV.}
\end{center}
\end{figure}

In Fig. 1 the spin-flip M1 strength (\ref{eq:strength_function}), calculated
with $g$-factors $g^{p}_s = 5.58 \varsigma_p$ and $g^{n}_s = - 3.82
\varsigma_n$, is presented in the deformed $^{158}$Gd and spherical $^{208}$Pb.
Both SRPA and unperturbed
strengths are shown to demonstrate the collective shift caused by the
residual interaction.  The forces SkO, SG2 and SV-bas are used as
representative examples.  The results are compared with the
experimental data which indicate a two-peak structure of the M1GR in
$^{158}$Gd and one (isovector) peak in $^{208}$Pb.

The figure illustrates a typical situation already pointed out in the study
\cite{vesely_PRC_09}, namely, that none of the considered Skyrme forces is able
to describe the M1 strength simultaneously in both deformed and doubly-magic
nuclei.  Indeed, we see that SkO well reproduces the one-peak structure of the
M1GR in $^{208}$Pb but fails in offering two peaks in $^{158}$Gd. Vise versa,
SG2 and especially SV-bas succeed in the two-peak structure in $^{158}$Gd but
deliver a wrong resonance shape in $^{208}$Pb.

These results may be understood in terms of two key factors: i) the proton and
neutrons spin-orbit splittings, $E_\mathrm{so}^p$ and $E_\mathrm{so}^n$, which set the proton
and neutron branches of the unperturbed resonance, and ii) the residual
interaction which produces a collective shift $E_\mathrm{coll}$ (defined as a
difference between SRPA and unperturbed resonance centroids). Fig. 1 shows
that, for the forces SG2 and SV-bas, the proton and neutron unperturbed
branches appear as separated peaks in $^{208}$Pb and as one single peak with a
right shoulder in $^{158}$Gd. In both cases the proton low-energy peak is
higher since $g^{p}_s > g^{n}_s$. The residual interaction upshifts the
strength by 1-2 MeV, redistributes it in favor of the upper peak, and somewhat
enlarges the splitting. As a result, a distinctive two-peak structure is
formed. Instead for SkO, the relative spin-orbit splitting
$E_\mathrm{so}=E_\mathrm{so}^n-E_\mathrm{so}^p$ is very small and the proton
and neutron branches actually form one peak which is then upshifted by the
residual interaction.

This analysis illustrates the well known fact
\cite{Harakeh_book_01,Speth_91,Osterfeld_92,vesely_PRC_09,Sarriguren_M1}
that the quality of the description of the M1GR is mainly determined
by the ratio $E_\mathrm{coll}/E_\mathrm{so}$ between the collective
shift and relative spin-orbit splitting. If the initial
$E_\mathrm{so}$ is large, then a strong residual interaction with
$E_\mathrm{coll} > E_\mathrm{so}$ is necessary to mix the proton and
neutron branches, redistribute the strength to a higher energy, and
thus produce a one-peak resonance. Otherwise, a two-peak structure
persists. If instead $E_\mathrm{so}$ is small, then the unperturbed
resonance already has one peak which is then merely upshifted by the
residual interaction.
\begin{figure}
\begin{center}
\includegraphics[height=8cm,width=6cm,angle=-90]{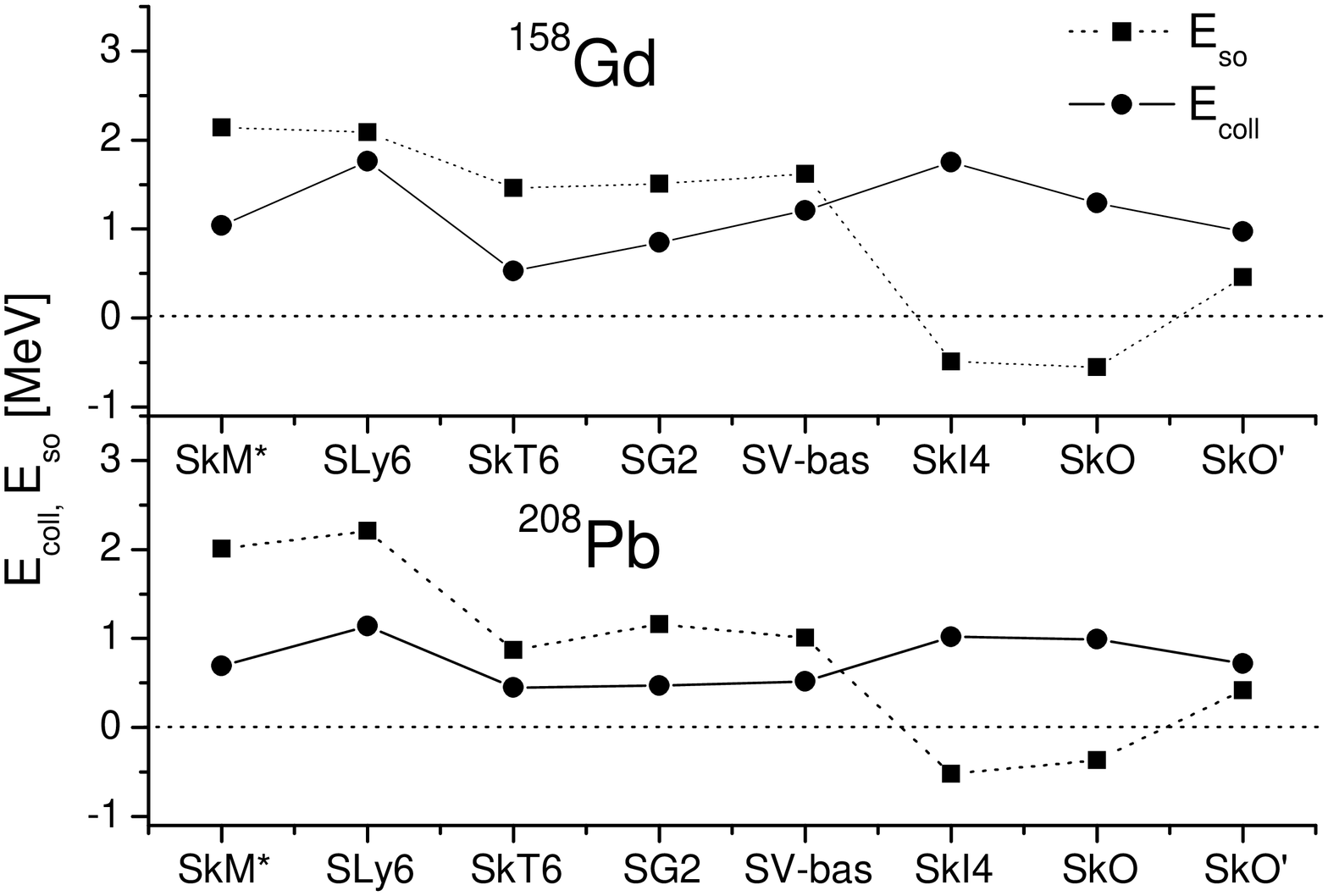}
\caption{ \label{fig:fig2_Gd_Pb_so_coll} The relative spin-orbit splittings
$E_{so}=E_{so}^n-E_{so}^p$ (full squares) and M1 collective shifts $E_{coll}$
(full circles) in $^{158}$Gd and $^{208}$Pb for 8 Skyrme forces. For better
view, the symbols are connected by lines. The horizontal line E=0 is drawn for
convenience of the comparison.}
\end{center}
\end{figure}

In Fig. 2, the key ingredients of the M1GR description,
$E_\mathrm{coll}$ and $E_\mathrm{so}$, are compared for eight Skyrme
forces. One sees that $E_\mathrm{coll} < E_\mathrm{so}$ for most of the
forces (SkM*, SLy6, SkT6, SG2, SV-bas) which should result in a
double-peak M1GR. And indeed, Fig. 1 demonstrate this for SG2 and SV-bas
in $^{158}$Gd and $^{208}$Pb. Instead, for the forces SkI4, SkO, and
SkO', we have $E_\mathrm{coll} > E_\mathrm{so}$ and hence the one-peak
M1GR. The results of Fig. 2 for $^{158}$Gd and $^{208}$Pb
remind those for $^{238}$U \cite{vesely_PRC_09}.  So the
similar results may be expected for other medium and heavy nuclei as
well.  This means that the M1GR structure is mainly determined
by the Skyrme force rather than by the particular nucleus. In other
words, the forces of the first (second) group should always yield a
two-peak (one-peak) structure. Hence a failure in simultaneous description of M1
strength in nuclei like $^{158}$Gd and $^{208}$Pb by one and the same force.
This is a very serious drawback of the
present-day Skyrme parameterizations.  Besides, this is an alarming
message for the SHF description of the GT resonance which, being a
counterpart of M1GR, is determined by the same factors.  Possible
ways to cure this problem will be discussed in the next sections.

\section{Isovector spin-flip M1(T=1) response}

In the above discussion and Ref. \cite{vesely_PRC_09}, we analyzed the spin-flip
M1 strength including both isovector (T=1) and isoscalar (T=0)
contributions. This strength was calculated with
$g$-factors $g^{p}_s=5.58\varsigma_p$ and $g^{n}_s=-3.82\varsigma_n$. As
was mentioned in Sec. \ref{sec:model}, the isovector $g$-factor is much
larger than the isoscalar one, $g^{1}_s >> g^{0}_s$, and so the M1
strength should be predominantly isovector.  In other
words, the M1 and purely isovector M1(T=1) responses are to be about the
same.  However, these arguments consider the M1GR as one entity and do
not take into account possible {\it local} differences (i.e. features
at particular energies) in M1 and M1(T=1) strengths. As is shown
below, these local differences can be essential and considerably
change the appearance of the M1GR.

In this connection, it is worth to compare with experiment the spin-flip
M1(T=1) response calculated with $g^{0}_s=0$ (similar calculations were
recently performed for $^{208}$Pb in \cite{Sagawa_lanl_09}). This differs from
most of the previous calculations
\cite{Kam_83,Pon_87,Zaw_90,Coster_90,Sarriguren_M1} where the common M1
strength was considered. However, the isovector separation is reasonable
because the experiment \cite{exp_Gd_U,exp_208Pb} treats the M1GR as the
isovector mode.

\begin{figure*}
\begin{center}
\includegraphics[height=10.5cm,width=8cm,angle=-90]{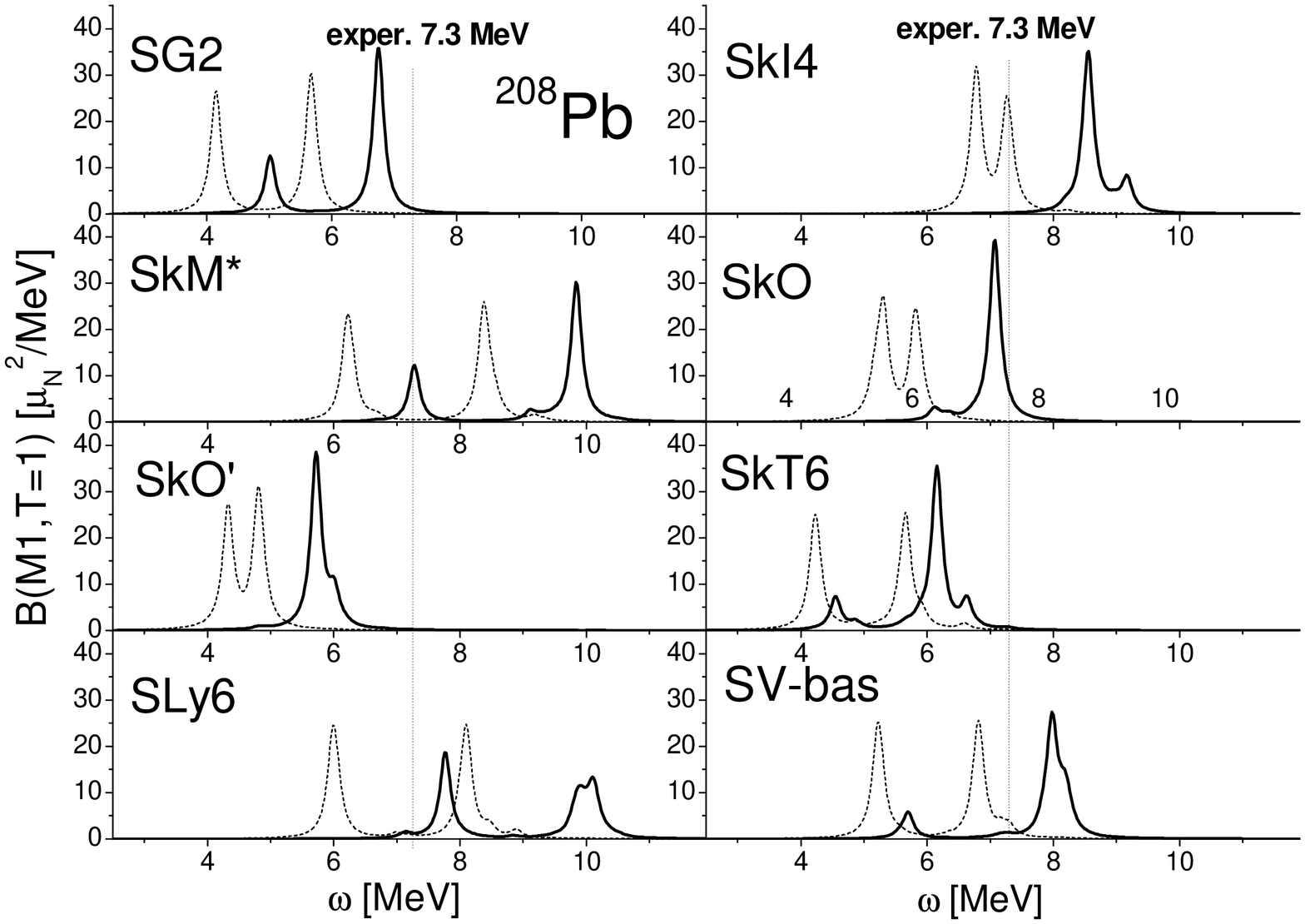}
\caption{\label{fig:fig1_T1Pb} Isovector M1(T=1) strength in $^{208}$Pb,
calculated with 8 Skyrme forces as indicated. The 2qp (short-dash curve) and
SRPA (solid curve) results are presented.  The vertical dash line marks the
average experimental resonance energy 7.3 MeV. The strength is smoothed by the
Lorentz weight with $\Delta$=0.2 MeV.}
\end{center}
\end{figure*}

In Fig. \ref{fig:fig1_T1Pb}, the isovector M1(T=1) strength in
$^{208}$Pb computed with eight different Skyrme forces is
presented. To discriminate the details, a small width of $\Delta$=0.2
MeV is used in the Lorentz smoothing. In this doubly-magic nucleus the
RPA spectrum is dilute and so the small smoothing does not cause an
excessive complication of the strength. Fig. \ref{fig:fig1_T1Pb} shows
that the results depend strongly on the force. However, unlike the M1 case,
the M1(T=1) strength already exhibits mainly a one-peak structure
provided by the dominant right peak. This structure is obvious even
for the forces SG2, SkM*, SkT6, and SV-bas, which show two peaks in M1
strength. Only SLy6 maintains the structure of the M1 result.
\begin{figure*}
\begin{center}
\includegraphics[height=9cm,width=9cm]{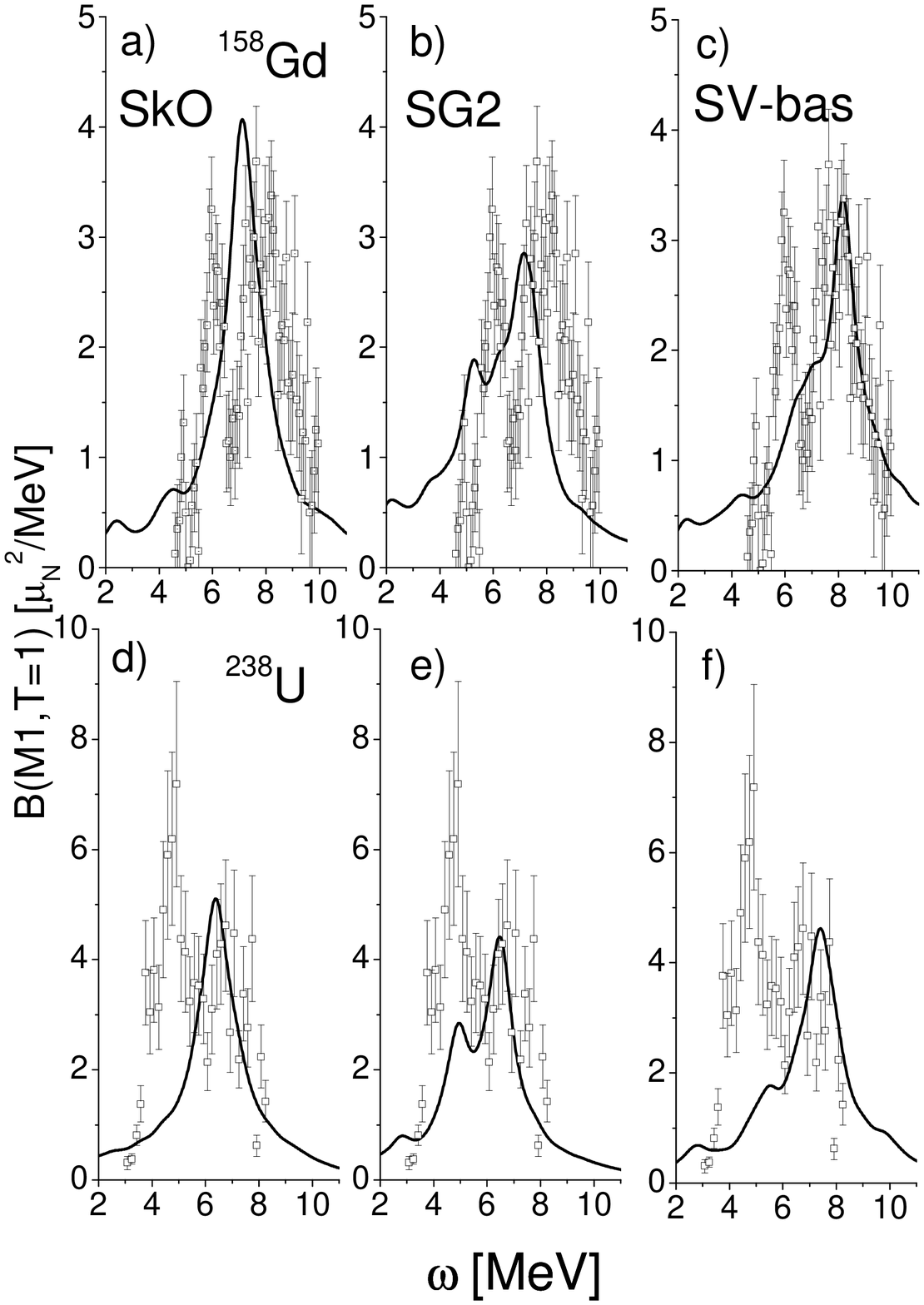}
\caption{ \label{fig:fig4_gd_U_T1} The isovector M1(T=1) strength in $^{158}$Gd
and $^{208}$Pb for the forces SkO (left), SG2 (middle), and SV-bas (right). The
experimental data \protect\cite{exp_Gd_U} are given by the grey boxes and bars.
The strength is smoothed by the Lorentz weight with $\Delta$=1 MeV.}
\end{center}
\end{figure*}

Such a difference between M1 and M1(T=1) responses may be explained in terms
of spin $g$-factors. The M1
transitions deal with $g^{p}_s = 5.58 \varsigma_p$ and $g^{n}_s = - 3.82
\varsigma_n$ and so, as was mentioned above, the unperturbed M1 strength
exhibits the left proton peak $\propto (g^{p}_s)^2$ about twice higher than the
right neutron peak $\propto (g^{n}_s)^2$. The residual interaction recasts the
M1 strength in favor of the right peak, which finally yields the two-peak
structure with comparable peak heights.  Instead, in M1(T=1) transitions we use
$g^{p}_s=-g^{n}_s=g^1_s/2$ = 3.12 and so, unlike the M1 case, the unperturbed
proton and neutron peaks already have about the same heights (with a bit higher
neutron peak). Then the further collective upshift of the strength results in
a strict dominance (more than in M1 case) of the right peak, hence mainly a
one-peak structure.
\begin{figure*}
\begin{center}
\includegraphics[height=11cm,width=8cm,angle=-90]{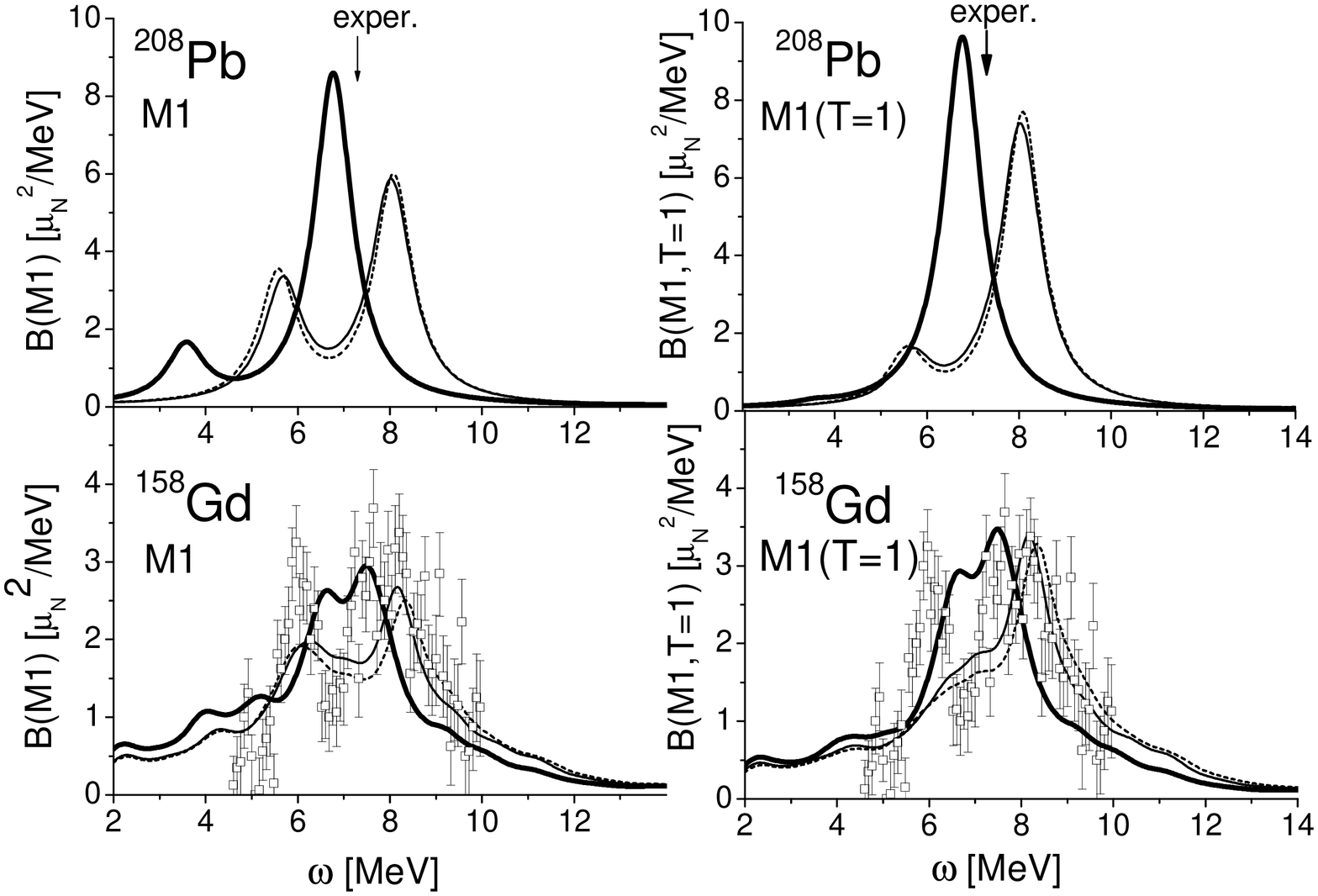}
\caption{ \label{fig:fig5_tensor} The spin-flip M1 (left) and M1(T=1) (right)
strengths in $^{208}$Pb and $^{158}$Gd for the force SV-bas in the standard
version (solid curve), with tensor contribution (bold curve) and  with
$b'_4=b_4$ (short-dash curve). The experimental data are given by boxes with
bars for $^{158}$Gd \protect\cite{exp_Gd_U} and vertical arrows for $^{208}$Pb
\protect\cite{exp_208Pb}. The strength is smoothed by the Lorentz weight with
$\Delta$=1 MeV. }
\end{center}
\end{figure*}

As a result, some forces provide an acceptable description of the
experimental data for the M1(T=1) case. The forces SG2, SkO, and
SV-bas give a dominant peak at the energies 6.8, 7.2, and 8 MeV,
i.e. close to the experimental average value 7.3 MeV. For other forces
the disagreement is larger than 1 MeV.

However, as already observed in case of mere M1 strength, an
acceptable agreement for spherical nuclei does not mean the same for
deformed ones.  Fig. 4 for $^{158}$Gd and $^{238}$U shows that SkO
does not reproduce the two-peak structure at all while SG2 and SV-bas
suggest it, but with strongly attenuated left peak. Though in
general the M1(T=1) response better agrees with the experiment for
$^{208}$Pb than the M1 response, a simultaneous description of the
experimental data in spherical and deformed nuclei still fails.

The comparison of Figs. 1 and 4 shows that for SV-bas the computed M1 strength
is closer to the experiment than the M1(T=1) one.
Then the natural question arises, to which extent the
experimental data \cite{exp_Gd_U} from $(p,p')$ reaction and \cite{exp_208Pb}
from $(\gamma,\gamma')$ reaction with tagged photons give just the isovector
M1GR?  Both experimental studies claim this. However this claim,
being based on the general reaction conditions, is actually not supported by
thorough estimations and checks. Moreover, these reactions should actually involve both
T=0 and T=1 channels. Thus the computation of the reaction cross-sections are
called for more adequate comparison with the experimental.
This uncertainty also could be one of the reasons of the disagreement between SHF and
experimental results for M1GR.

\section{Tensor and isovector spin-orbital forces}

Another point to be discussed in connection with the M1GR problems is
the influence of the tensor and T=1 spin-orbit interactions.  Both
interactions come to the Skyrme functional through the terms with
spin-orbit densities \cite{Ben}. As shown in our recent study
\cite{vesely_PRC_09}, these interactions can affect the M1GR through the spin-orbit
splittings $E^{p}_\mathrm{so}$ and $E^{n}_\mathrm{so}$. The tensor interaction
changes $E^{p}_\mathrm{so}$ and $E^{n}_\mathrm{so}$ likewise, thus producing a
total M1GR shift without a noticeable variation of the relative splitting
$E_\mathrm{so}$. Instead the T=1 spin-orbit interaction changes $E^p_\mathrm{so}$
and $E^n_\mathrm{so}$ on scale and so affects $E_\mathrm{so}$.

Here we take into account both static mean-filed and collective impacts
of the tensor and T=1 spin-orbit interactions. The results are demonstrated
in Fig. 5. It is worth reminding that the
spin-orbit term in Skyrme functional (\ref{eq:skyrme_funct}) reads
\begin{equation}
 - b_4 (\rho \nabla\textbf{J}\!+
        \!(\nabla\!\times\!\textbf{j})\!\cdot\!\textbf{s})
 - b'_4 \sum_q (\rho_q \nabla\textbf{J}_q\!
 +\!(\nabla\!\times\!\textbf{j}_q)\!\cdot \!\textbf{s}_q) .
\end{equation}
The standard SHF calculations use $b'_4=b_4$, i.e. only the isoscalar (T=0)
contribution. At the same time, the relativistic models employ $b'_4$=0
\cite{Ben,Vre05aR,Sto07aR}. So, it is worth to decouple the coefficients
$b'_4$, and $b_4$ and thus introduce the isovector (T=1) spin-orbit
interaction, as was done e.g. in \cite{ski3}. Such decoupling is natural
since a similar separation is already used for other Skyrme coefficients,
$b_i$ and $b'_i$, with
$i=0,1,2,3$. Actually, the force SV-bas already follows this track and uses
$b_4=34.117$ and $b'_4 = 0.547b_4$. The effect is demonstrated in Fig. 5 for M1
and M1(T=1) responses, where the SV-bas result is compared with $b'_4 = b_4$
variant (after refitting). It is seen that the effect is not large. However,
any final conclusions on its scale can be done only after thorough checks
involving various Skyrme forces and nuclei.

We now consider the impact of tensor interaction. It is often omitted in the
standard effective two-body Skyrme interaction. If included, it adds to the
functional (\ref{eq:skyrme_funct}) the term
\begin{equation}\label{tensor}
  \gamma_\mathrm{T}(\tilde{b}_1
   (\textbf{s}\!\cdot\!\textbf{T}\!-\!\textbf{J}^2)
  + \tilde{b}'_1
   \sum_q (\textbf{s}_q\!\cdot\!\textbf{T}_q
    \!-\!\textbf{J}_q^2))
\end{equation}
where the squared spin-orbit densities $\textbf{J}^2$ and $\textbf{J}^2_q$
represent the tensor contribution while $\textbf{s}\!\cdot\!\textbf{T}$ and
$\textbf{s}_q\!\cdot\!\textbf{T}_q$ terms serve to restore in (\ref{tensor})
the Galilean invariance.  The exchange part of the zero-range Skyrme
interaction also leads to similar spin-orbit terms, see e.g.
\cite{Colo_Sagawa_08}.  To be accurate, the tensor and central exchange
contributions should be treated separately and their parameters are to be
determined from the initial effective two-body interaction
\cite{Colo_Sagawa_08,Sagawa_lanl_09}. However, from the point of view of a zero-range Skyrme
interaction, it is reasonable not to distinguish the tensor and central
exchange terms and use for both of them the same fitting parameters
$\tilde{b}_1$ and $\tilde{b}'_1$. We use here just such common practice. For
simplicity, the tensor and central exchange contributions will be further
called tensor terms. Note that these tensor terms influence both ground state
properties and dynamics.
In the results shown in Figs. 1-4, the forces SkO and SV-bas have no
tensor terms. However, these terms are added in SG2 as they
noticeably improve description of the ground state for this particular
parameterization.

As an example, we will now compare the SV-bas results with and without tensor
terms. They are fully switched on by $\gamma_\mathrm{T}=1$.
As shown in \cite{vesely_PRC_09}, the refitting of other Skyrme parameters
may considerably decrease the tensor effect. So, we use for $\gamma_\mathrm{T}=1$
the refitted SV-bas parameters. The tensor contributions to both ground
state and SRPA residual interaction are taken into account.
The results of the calculations are shown in Fig. 5.
It is seen that the tensor effect is indeed dramatic (a large tensor impact
on M1GR was also found in \cite{Sagawa_lanl_09}).
Moreover, it considerably improves agreement with the
experimental data in this particular case. Hence the tensor forces
can indeed be an important factor in the description of M1GR.

It is also worth noting that tensor forces significantly influence the
Landau-Migdal parameters $g_0$ and $g'_0$ in the spin and spin-isospin
channels and affect the estimation of spin instability of nuclear
matter for Skyrme forces \cite{Marg_Sagawa_JPG_09,Bender_PRC_02_GT}.
Eight Skyrme parameterizations used in the present study have $g'_0>-1$
and so are spin-stable at the equilibrium density in the spin-isospin
channel. Just this channel determines the isovector spin-flip M1 and
Gamow-Teller GR.  Anyway, our knowledge on the interplay between tensor forces
and spin correlations is still rather poor and M1GR could be used here as a
robust and important test in clarification of this interplay.

\section{Conclusions}

The open problem of the description of the spin-flip M1 giant
resonance (M1GR) within the Skyrme-Hartree-Fock (SHF) approach is
analyzed. It is shown that presently available Skyrme
parameterizations poorly reproduce the experimental data and, in
particular, cannot provide a simultaneous description of M1GR gross
structure in deformed and spherical (doubly magic) nuclei. The two
main factors responsible for the M1GR properties, spin-orbit
splittings and spin correlations, are inspected for eight different
Skyrme parameterizations.

Some critical aspects are worked out. One point is the essential
difference between M1 and M1(T=1) responses which leads to the open
question: how much is the observed strength of the isovector nature
and which of the responses should be compared with it? Furthermore, the
essential influence of the tensor force was demonstrated, which can have
really dramatic effects. So the tensor interaction can be a key element in
the further development of a better M1GR description. An appropriate
T=1 part of the spin-orbit interaction could also be an important
ingredient.

Altogether, the SHF description of the M1GR remains yet open as a
quite complicated problem where many contributions are
entangled. The problem may have general consequences for the SHF description of
nuclear dynamics in the spin-isospin channel. More development is
needed to establish SHF as a reliable model also for spin properties.

\begin{acknowledgments}
The work was partly supported by the DFG RE-322/12-1, Heisenberg-Landau
(Germany - BLTP JINR), and Votruba - Blokhintsev (Czech Republic - BLTP JINR)
grants. W.K. and P.-G.R. are grateful for the BMBF support under contracts 06
DD 9052D and 06 ER 9063. Being a part of the research plan MSM 0021620859
(Ministry of Education of the Czech Republic) this work was also funded by
Czech grant agency (grant No. 202/06/0363). P.V. is grateful for the FIDIPRO
support. V.Yu.P. thanks the DFG support under the Contract SFB 634.
\end{acknowledgments}



\begin{thebibliography}{99}
\bibitem{Harakeh_book_01} 
 Harakeh M N and van der Woude A 2001 {\it Giant Resonances}
 (Oxford: Clarendon Press)
\bibitem{Speth_91} 
  {\it Electric and magnetic giant resonances in nuclei} 1991
  ed. Speth J (Singapore: World Scientific)
\bibitem{Osterfeld_92} 
   Osterfeld F 1992 {\it Rev. Mod. Phys.} {\bf 64} 491
\bibitem{Kam_83} 
  Kamerdzhiev S P and Tkachev V N 1984
   {\it Phys. Lett.} B {\bf 142} 225
\bibitem{Pon_87} 
  Ponomarev V Yu, Vdovin A I and Stoyanov Ch 1987
  {\it J. Phys G: Nucl. Part. Phys.} {\bf 13} 1523
\bibitem{Zaw_90} 
  Zawischa D and Speth J 1990
  {\it Phys. Lett.} B {\bf 252} 4
\bibitem{Coster_90} 
  de Coster, Heide K and Richter A 1990
  {\it Nucl. Phys.} A {\bf 542} 375
\bibitem{Ben} 
  Bender M, Heenen P-H and Reinhard P-G 2003
  {\it Rev. Mod. Phys.} {\bf 75} 121
\bibitem{Vre05aR} 
 Vretenar D, Afanasjev A V, Lalazissis G A and Ring P 2005
 {\it Phys. Rep.} {\bf 409} 101
\bibitem{Sto07aR} 
 Stone J R and Reinhard P-G 2007
 {\it Prog. Part. Nucl. Phys.} {\bf 58} 587
\bibitem{Les_PRC_07_tensor} 
 Lesinski T, Bender M, Bennaceur K, Duguet T and Meyer J 2007
{\it Phys. Rev.} C {\bf 76} 014312
\bibitem{Colo_Sagawa_08} 
  Zow W, Colo G, Ma Zh, Sagawa H and Bortignon P F 2008
{\it Phys. Rev.} C {\bf 77}, 014314
\bibitem{Marg_Sagawa_JPG_09} 
 Margueron J and Sagawa H 2009
  J. Phys. G: Nucl. Part. Phys. {\bf 36} 125102;
 Margueron J, Goriely S, Grasso M and Sagawa H 2009
ibid {\bf 36} 125103
\bibitem{Sagawa_lanl_09} 
  Cao Li-Gang, Colo G, Sagawa H, Bortignon P F and Sciacchitano L 2009
arXiv:0909.4433[nucl-th].
\bibitem{vesely_PRC_09} 
 Vesely P, Kvasil J, Nesterenko V O, Kleinig W, Reinhard P-G
 and Ponomarev V Yu 2009 {\it Phys. Rev.} C {\bf 80} 031302(R)
\bibitem{Skyrme} 
  Skyrme T H R 1956
  {\it Phil. Mag.} {\bf 1} 1043
\bibitem{Vau} 
  Vautherin D and Brink D M 1972
  {\it Phys. Rev.} C {\bf 5} 626
\bibitem{Engel_75} 
  Engel Y M, Brink D M, Goeke K, Krieger S J and Vauterin D 1975
  {\it Nucl. Phys.} A {\bf 249} 215
\bibitem{Sarriguren_M1} 
 Sarriguren P, Moya~de~Guerra E and Nojarov R 1996
 {\it Phys. Rev.} C {\bf 54} 690
\bibitem{Hilton_98} 
  Hilton R R, H$\ddot o$henberger W and Ring P 1998
  {\it Eur. Phys. J.} A {\bf 1} 257
\bibitem{nest_IJMPE_09} 
 Nesterenko V O, Kvasil J, Vesely P, Kleinig W and Reinhard P-G,
 to be published in {\it Int. J. Mod. Phys.} E; arXiv:0911.2410[nucl-th].
\bibitem{nest_PRC_02} 
  Nesterenko V O, Kvasil J and Reinhard P-G 2002
  {\it Phys. Rev.} C {\bf 66} 044307
\bibitem{nest_PRC_06} 
  Nesterenko V O, Kleinig W, Kvasil J, Vesely P, Reinhard P-G
  and Dolci D S 2006
 {\it Phys. Rev.} C {\bf 74} 064306
\bibitem{nest_PRC_08} 
  Kleinig W, Nesterenko V O, Kvasil J, Reinhard P-G and Vesely P 2008
  {\it Phys. Rev.} C {\bf 78} 044313
\bibitem{Petr_PhD} 
  Vesely P 2009 Ph. D. thesis, Charles University in Prague, Czech Rep.
\bibitem{exp_Gd_U} 
 Frekers D et al 1990 {\it Phys. Lett.} B {\bf 244} 178;
 W$\ddot o$rtche H L 1994 Ph.D. thesis,
  Technischen Hochschule Darmstadt, Germany.
\bibitem{exp_208Pb} 
 Laszewski R M, Alarcon R, Dale D S and Hoblit S D 1988
 {\it Phys. Rev. Lett.} {\bf 61} 1710
\bibitem{skt6} 
  Tondeur F, Brack M, Farine M and Pearson J M 1984
  {\it Nucl. Phys.} A {\bf 420} 297
\bibitem{sko} 
 Reinhard P-G, Dean D J, Nazarewicz W, Dobaczewski J,
 Maruhn J A and Strayer M R 1999
 {\it Phys. Rev.} C {\bf 60} 014316
\bibitem{sg2} 
  Van Giai N and Sagawa H 1981 {\it Phys. Lett.} B {\bf 106} 379
\bibitem{skms} 
  Bartel J, Quentin P, Brack M, Guet C and H\aa{a}kansson H-B 1982
  {\it Nucl. Phys.} A {\bf 386} 79
\bibitem{sly46} 
  Chabanat E, Bonche P, Haensel P, Meyer J and Schaeffer R 1997
  {\it Nucl. Phys.} A {\bf 627} 710
\bibitem{ski3} 
  Reinhard P-G and Flocard H 1995
  {\it Nucl. Phys.} A {\bf 584} 467
\bibitem{svbas} 
 Kl\"{u}pfel P, Reinhard P-G, B\"{u}rvenich T J and Maruhn J A 2009
 {\it Phys. Rev.} C {\bf 79} 034310
\bibitem{nest_ijmp} 
   Nesterenko V O, Kleinig W, Kvasil J, Vesely P and Reinhard P-G 2007
    {\it Int. J. Mod. Phys.} E {\bf 16} 624; 2008 ibid {\bf 17} 89
\bibitem{Bender_PRC_02_GT}
  Bender M, Dobaczewski J, Engel J and Nazarewicz W 2002
  {\it Phys. Rev.} C {\bf 65} 054322
\bibitem{Fracasso_PRC_07_GT} 
 Fracasso S and Colo G 2007
{\it Phys. Rev.} C {\bf 76} 044307;
 Bai C L et al 2009 {\it Phys. Lett.} B {\bf 675} 28
\bibitem{Sarrin_NPA_01_GT} 
 Sarriguren P, Moya de Guerra E and Escuderos A 2001
 {\it Nucl. Phys.} A {\bf 691} 631
\end{thebibliography}
\end{document}